\documentclass{article}
\usepackage{spconf}
\usepackage{tabu}

\usepackage[]{siunitx}

\usepackage{cite}
\usepackage{graphicx}
\usepackage[cmex10]{amsmath}
\usepackage{multirow}
\usepackage{makecell}
\usepackage{diagbox}

\usepackage{newtxtext,newtxmath}
\usepackage{bm}
\usepackage{psfrag}
\usepackage{color}
\usepackage{colortbl}
\usepackage{booktabs}

\usepackage{arydshln}

\title{Profile-error-tolerant target-speaker voice activity detection}

\makeatletter
\def\@name{\textit{Dongmei Wang, Xiong Xiao, Naoyuki Kanda, Midia Yousefi, Takuya Yoshioka, Jian Wu}\\ }
\makeatother
\address{Microsoft, One Microsoft Way, Redmond, WA, USA}

\begin{document}
\ninept

\maketitle

\begin{abstract}

Target-Speaker Voice Activity Detection (TS-VAD) utilizes a set of speaker profiles alongside an input audio signal to perform speaker diarization. While its superiority over conventional methods has been demonstrated, the method can suffer from errors in speaker profiles, as those profiles are typically obtained by running a traditional clustering-based diarization method over the input signal. This paper proposes an extension to TS-VAD, called Profile-Error-Tolerant TS-VAD (PET-TSVAD), which is robust to such speaker profile errors. This is achieved by employing transformer-based TS-VAD that can handle a variable number of speakers and further introducing a set of additional pseudo-speaker profiles to handle speakers undetected during the first pass diarization. During training, we use speaker profiles estimated by multiple different clustering algorithms to reduce the mismatch between the training and testing conditions regarding speaker profiles. Experimental results show that PET-TSVAD consistently outperforms the existing TS-VAD method on both the VoxConverse and DIHARD-I datasets.

\end{abstract}


\vspace{-.5em}
\section{Introduction}
\label{sec:Introduction}
\vspace{-.5em}

Speaker diarization is a technique to recognize ``who spoke when'' for multi-talker conversation recordings \cite{park_diarization_review_2013}. It is an essential component for many speech-related applications, such as automatic meeting transcription, meeting summarization, and call center data analytics. 
Traditional speaker diarization systems consist of voice activity detection (VAD), segmentation, speaker embedding extraction, and clustering modules \cite{kmeans_2013, ahc_2014}. While such a clustering-based approach achieved promising results \cite{ivector_2018, diez2020dihard}, it cannot deal with overlapping speech, which is often observed in natural conversation \cite{shriberg2001observations}. 
Although several extensions to combine overlapping speech detection have been proposed \cite{bullock2020overlap,wang2021bytedance}, such systems tend to be complicated and require a careful tuning of the hyperparameters.

Recently, end-to-end neural speaker diarization (EEND) \cite{fujita_eend_interspeech_2019, fujita_eend_asru_2019} was proposed to handle overlapping speech. The EEND model is trained by permutation invariant training (PIT) to directly output multiple speaker activities. To handle a variable number of speakers, an EEND with encoder-decoder-attractor (EEND-EDA) is later proposed \cite{shota_eendEDA_2020, shota_eendEDA_2022}.
In another line of research to deal with overlapping speech,
TS-VAD \cite{medennikov2020stc,ivan_TSVAD_2020} was proposed with 
great success in diarization in adverse acoustic environments with large portion of speech overlaps \cite{watanabe20b_chime}.
TS-VAD model use a set of speaker profiles alongside an input audio signal to estimate the voice activities of each speaker. 
The original TS-VAD model can only deal with a fixed number of speakers.
This issue was initially addressed by using heuristics during the inference  \cite{duke_VoxSRC_2021, Wang2021DirHardIII}. 
Later, several extensions to the model architecture were proposed to handle variable number of speakers, including transformer-based TS-VAD \cite{dwang2023_icassp_edaTSvad} and multi-target filter and detector
\cite{ chinyi_multiTargetFilter_2022}. 

While TS-VAD models achieved strong speaker diarization accuracy
in many scenarios \cite{watanabe20b_chime, Wang2021DirHardIII, duke_VoxSRC_2021, brown2022voxsrc, cheng2022seq2seqtsvad},
they can suffer from errors in speaker profiles, 
as those profiles are typically obtained by running a traditional
clustering-based diarization method over the input signal. 
For example, if some speakers are not detected in the initial speaker diarization,
the current TS-VAD model cannot detect such speakers.
In another example, if the speeches from the same speaker 
are classified into multiple clusters,
the current TS-VAD model will detect multiple speaker activities 
based on the subtle differences of speaker profiles from such clusters.

In this work, we propose a Profile-Error-Tolerant TS-VAD (PET-TSVAD), a novel TS-VAD that is tolerant to the speaker profile errors. 
Our proposal is based on the transformer-based TS-VAD model that can handle a variable number of speakers \cite{dwang2023_icassp_edaTSvad}.
To address the case that some speakers are not detected in the initial diarization, we introduce a set of pseudo-speaker profiles and concatenate them with the detected speaker profiles as the input of the TS-VAD. 
During the training, the pseudo-speaker profiles are trained to estimate the undetected speakers' speech activities.  Additionally, instead of using oracle speaker profiles for training,
we propose to use speaker profiles estimated by multiple different clustering algorithms to reduce the mismatch between the training and testing conditions.
We evaluate the proposed PET-TSVAD model by using
VoxConverse \cite{Voxconverse2021dataset} and DIHARD-I \cite{dihard12018} datasets,
and show its superiority over the conventional TS-VAD models.

\vspace{-.5em}
\section{Transformer-based TS-VAD: Review}
\label{sec:tsvad_review}
\vspace{-.5em}

The transformer-based TS-VAD \cite{dwang2023_icassp_edaTSvad} was proposed
as an extension of TS-VAD \cite{medennikov2020stc,ivan_TSVAD_2020}
to deal with a variable number of speakers for diarization.
Similar to the original TS-VAD, 
the input to the transformer-based TS-VAD model includes log mel-filterbank features and the speaker profiles. 
The model then outputs the frame-wise speech activities for all input speaker profiles. 

The transformer-based TS-VAD \cite{dwang2023_icassp_edaTSvad} consists of the encoder module, independent speaker detection (ISD) module, and joint speaker detection (JSD) module. 
The encoder module first converts the log mel-filterbank features into frame-level embeddings $\textbf{E}\in\mathbb{R}^{T\times E}$, where $T$ and $E$ are the sequence length and dimension of the embeddings, respectively. 
The ISD module then processes each target speaker 
independently based on their profiles $\textbf{P}=[\textbf{p}_1,...,\textbf{p}_S]^{\textsc{T}}\in\mathbb{R}^{S\times P}$, 
 where $S$ is the number of the target speakers, 
$P$ is the profile dimension, and $^{\textsc{T}}$ denotes matrix transpose. 
Specifically, for the $i$-th target speaker, 
the input matrix
of shape $\mathbb{R}^{T\times (E+P)}$ is formed
by appending its profile vector $\textbf{p}_i$
to all the $T$ embedding vectors in $\textbf{E}$. 
The ISD module consists of a linear projection layer, followed by two
 bidirectional long short-term memory (BLSTM) layers. 
The output from the ISD module is $S$ matrices of
shape $\mathbb{R}^{T\times F}$,
where $F$ is the output dimension of the BLSTM layer.
The JSD module 
then treats the $S$ matrices as a 3D tensor with a shape of $\mathbb{R}^{S \times T \times F}$, and applies sequence modeling layers for 
the time-axis and the speaker-axis in alternating order.
Specifically, for the time-axis, the BLSTM layer is applied for modeling the temporal correlation. 
For the speaker-axis, the transformer layer without positional encoding is applied to model the cross-speaker correlation, making the model agnostic to the number and order of the input speaker profiles. 
Such a process is repeated twice to model the temporal and speaker correlation for the input speech signal. 
Finally, a linear layer with sigmoid activation is applied to obtain the speech activities for all the target speakers. 
Please refer to \cite{dwang2023_icassp_edaTSvad} for more details.

\vspace{-.5em}
\section{PET-TSVAD}
\vspace{-.5em}

\subsection{Motivation}
\vspace{-.5em}

The TS-VAD models (both the original TS-VAD and the transformer-based TS-VAD)
are typically applied after a traditional clustering-based speaker diarization method
as they require the speaker profiles as an input.
While TS-VAD models achieved strong speaker diarization performance, 
they can suffer from errors in such a first pass diarization.
There are two typical errors in the first pass diarization, i.e. {\bf merging} multiple speakers into one cluster and {\bf splitting} a single speaker into multiple clusters. 
These can happen simultaneously and make the speaker profiles contaminated.

Existing TS-VAD models cannot handle such speaker profile errors because of the structural limitation and training data mismatches.
First, the existing TS-VAD models are designed under an assumption of a one-to-one mapping between the input profiles and output speaker activities.
Therefore, if the ``merging'' type error happens in speaker profiles,
the existing TS-VAD models cannot detect most of speakers merged into one cluster.
Second, the existing TS-VAD models are 
trained with oracle speaker profiles, where the profiles are both complete (i.e., no missing profiles) and clean (i.e., each profile represents only one speaker's characteristics). 
Therefore, if the ``splitting'' type error occurs in speaker profiles, 
the TS-VAD models attempt to predict multiple speech activities based on the subtle differences of the speaker profiles.

In this paper, we propose PET-TSVAD,
which is tolerant to such speaker profile errors.
In the following subsections, we explain
the modification to the model architecture (Section \ref{sec:model_arch}),
the speaker profile in training (Section \ref{sec:profile_generation_procedure}),
and how we train the PET-TSVAD model (Section \ref{sec:training}).

\vspace{-.5em}
\subsection{Model architecture}
\label{sec:model_arch}
\vspace{-.5em}

\begin{figure}[t]
    \centering
    \includegraphics[scale=0.36]{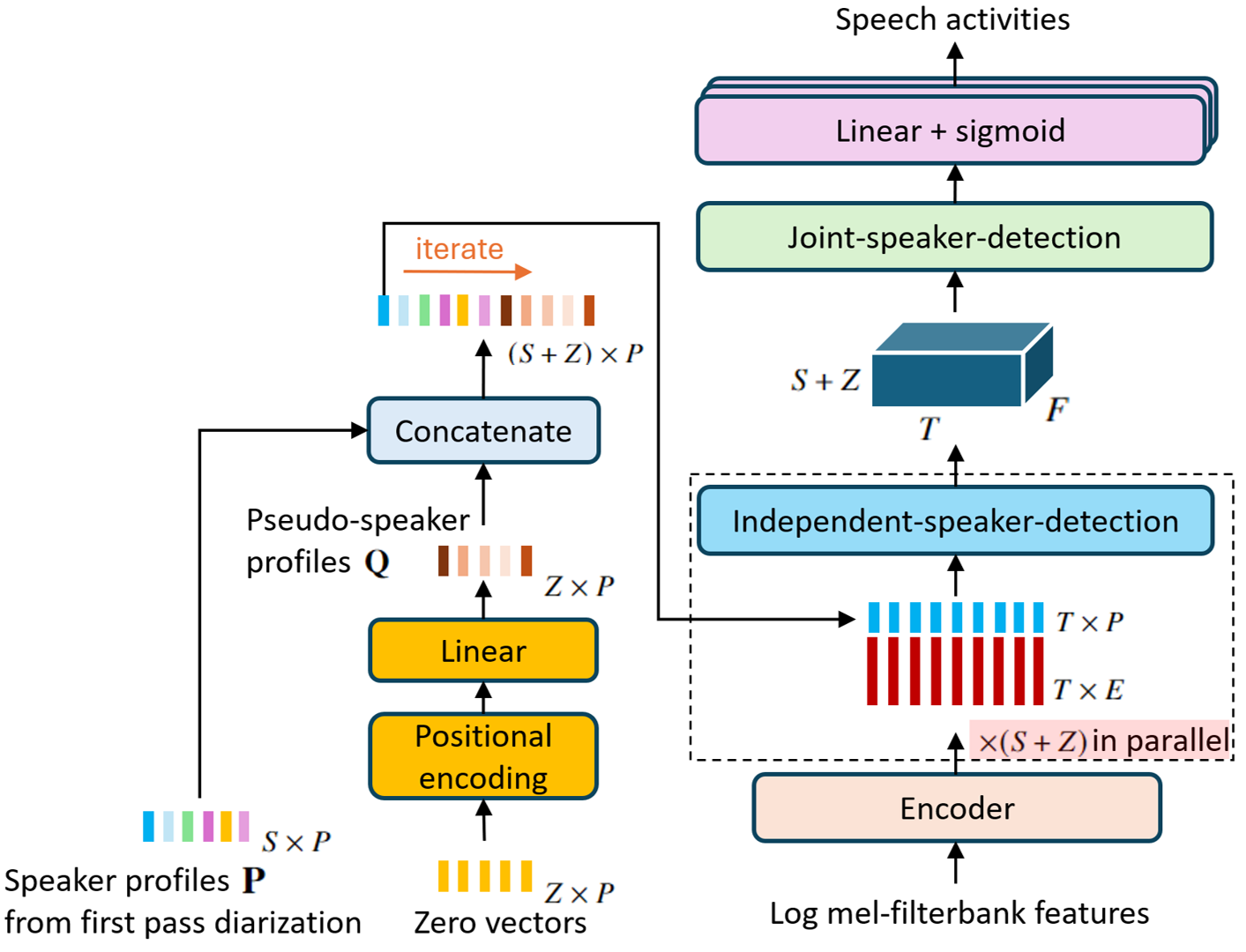}
    \vspace{-7mm}
    \caption{Model structure of PET-TSVAD.} 
    \label{fig: model_structure} 
     \vspace{-3mm}
\end{figure}

To address the structural limitation of the existing TS-VAD models, we propose to add a few pseudo-speaker profiles to augment the target speaker profiles from the first pass diarization. For those speakers without a valid profile from the first pass diarization, we expect their activities to be assigned in one of the pseudo-speaker profiles' output VAD dimension. Fig.~\ref{fig: model_structure} illustrates the model architecture of the PET-TSVAD. 
The profiles fed to the model are from two sources. The first source is the target speaker profiles generated from the first pass diarization $\textbf{P}\in\mathbb{R}^{S\times P}$.
The second source is a fixed number of learnable pseudo-speaker profiles $\textbf{Q}\in\mathbb{R}^{Z\times P}$, where $Z$ is the number of pseudo-speaker profiles. 
These pseudo-speaker profiles are introduced to detect missing speakers in the first pass diarization. 
They are derived from zero-vectors $\textbf{O}\in\mathbb{R}^{Z\times P}$ followed by positional encoding and a linear layer, where the positional encoding is used to ensure the pseudo-speaker profiles have different values. A linear layer is applied to transform the positional encoding into the speaker profile space. The pseudo-speaker profiles $\textbf{Q}$ are concatenated with the target speaker profiles $\textbf{P}$ to form the augmented speaker profiles with a shape of $\mathbb{R}^{(S+Z)\times P}$, which are fed to the TS-VAD module alongside the log mel-filterbank features.

To save memory and computation, PET-TSVAD is usually applied to short chunks of long-form audio independently during both training and the inference. When such a chunk-based processing is applied in inference, one speaker may appear at different pseudo-speaker positions across different chunks. We will investigate if it is necessary to perform the clustering on these pseudo-speaker embeddings across all chunks in our experiments.

\vspace{-.5em}
\subsection{Speaker profile generation for training}
\label{sec:profile_generation_procedure}
\vspace{-.5em}

In order to reduce the mismatch of speaker profile quality between training and the testing, we generate speaker profiles with various clustering algorithms for training the PET-TSVAD model. 
For each long-form audio in the training set, we apply the traditional clustering based diarization to obtain the cluster centers as target speaker profiles. To increase the diversity of the profile errors, we use two types of clustering, including the agglomerative hierarchical clustering (AHC)~\cite{han2007AHC} and the normalized maximum eigengap spectral clustering (NME-SC)~\cite{taejin_eigengap_2019}. For AHC, we used different stopping thresholds to further increase the data diversity. We also include a configuration that skips the VAD step to introduce more clustering errors. 
After clustering, we compute a target speaker profile for each cluster. The profile is computed as the average speaker embedding vectors over all the frames belonging to the cluster and a ResNet-based d-vector extractor \cite{zhou2019cnn} is used. During training, a longform audio is chunked into multiple training samples, and all of them share the same set of target speaker profiles.

\begin{table*}[t]
\centering
\caption{Data ratios of two configurations for speaker profile extraction during training}
\label{tab:data-ratio-cluster}
\footnotesize
  \vspace{0.5mm}
\scalebox{0.82}{
\begin{tabular}{@{}cccccccccccccccccccccccc@{}}
\toprule
\multirow{2}{*}{Clustering}      && \multicolumn{8}{c}{AHC w/o VAD}                           && \multicolumn{8}{c}{AHC w/ VAD}   && \multirow{2}{*}{\shortstack[l]{NME-SC\\w/o VAD}}      && \multirow{2}{*}{\shortstack[l]{NME-SC\\w/ VAD}}      &   \multirow{2}{*}{Oracle}                      \\ \cmidrule{3-10} \cmidrule{12-19}
      
 && 0.89 & 0.90  & 0.91 & 0.92 & 0.93 & 0.96 & 0.97 & 0.98 && 0.89 & 0.90  & 0.91 & 0.92 & 0.93 & 0.96 & 0.97 & 0.98 &&             &&            &                        \\ \midrule
Config. 1     &&   -   &  -    &  -    &   -   & -     & 0.25 & 0.25 & 0.25 &&    -  &  -    &  -    &   -   & - & - &  -    &  -    &      &  -    &              &    -         & 0.25                    \\
Config. 2     && 0.06 & 0.06 & 0.06 & 0.06 & 0.06 & 0.02 & 0.02 & 0.02 && 0.06 & 0.06 & 0.06 & 0.06 & 0.06 & 0.02 & 0.02 & 0.02 && 0.06         && 0.06        & 0.16                    \\ \bottomrule
\end{tabular}}
\end{table*}

\vspace{-.5em}
\subsection{Model training}
\label{sec:training}
\vspace{-.5em}

In previous TS-VAD models, there is a fixed one-to-one mapping between the estimated speaker activities and the reference speaker labels, as the oracle speaker profiles are used as input for training. 
Hence, the speaker-wise binary cross entropy is usually used as the training loss. 
However, in the proposed framework, the speaker profiles are obtained from clustering and may contain errors.
Therefore, a one-to-one mapping between the estimated speaker activities and the reference speaker labels is no longer guaranteed. In addition, the pseudo-speaker profiles introduce additional ambiguity of the output-to-reference mapping. 

To overcome this challenge, we adopt PIT \cite{Yu_pit_2017} for training the PET-TSVAD model, where we compute the binary cross entropy loss for all possible output-reference pairs and select the best permutation to perform the error backpropagation. For computational efficiency, we use the Hungarian algorithm \cite{Kuhn1955Hungarian} to find the best permutation.

\begin{table}[t]
\centering
\tabcolsep = 1.2mm
\caption{DER (\%) of the clustering algorithms on development sets. Different stopping thresholds are used for AHC.}
\label{tab:data-ratio-der}
\vspace{0.5mm}
\scalebox{0.81}{
\begin{tabular}{@{}ccccccccccc@{}}
\toprule
\multirow{2}{*}{Data} & \multicolumn{3}{c}{AHC w/o VAD} && \multicolumn{3}{c}{AHC w/ VAD} && \multicolumn{2}{c}{NME-SC}                                \\ \cmidrule{2-4} \cmidrule{6-8} \cmidrule{10-11}
                      & 0.90       & 0.92     & 0.97     && 0.90      & 0.92     & 0.97     && \multicolumn{1}{l}{w/o VAD} & \multicolumn{1}{l}{w/ VAD} \\ \midrule
VoxConverse                   & 8.94      & 10.9     & 26.26    && 7.19     & 8.63     & 24.25    && 10.88                       & 10.58                      \\
DIHARD-I                   & 47.18	&47.21	&51.02	&&41.75	&42.18	&45.28	&&47.76	&42.51   \\ \bottomrule
\end{tabular}}
\end{table}

\vspace{-.5em}
\section{Experiments}
\vspace{-.5em}

\subsection{Evaluation data}\label{eval_data}
\vspace{-.5em}

We conducted the experiments on VoxConverse \cite{Voxconverse2021dataset} and \mbox{DIHARD-I} \cite{dihard12018} data sets. 
The VoxConverse data set was developed based on YouTube videos for audio-visual diarization and used for the audio-based diarization track of VoxSRC 2020 \cite{nagrani2020voxsrc} and VoxSRC 2021 \cite{brown2022voxsrc} challenges. It consists of a development set and a test set. 
The development set contains 216 sessions with 20.3 hours of audio data. There are 1 to 20 speakers in each session of the development set. The test set contains 232 sessions with a total audio length of 43.5 hours. The speaker number for each session ranges from 1 to 21 in the test set. 
The diarization error rates (DER) was calculated with a 0.25-second of tolerance collar by following the description of the VoxConverse dataset~\cite{Voxconverse2021dataset}. 

The DIHARD-I data set was created for the first DIHARD challenge \cite{dihard12018}. It contains a diverse range of datasets, including clinical interviews, restaurant conversation, meeting speech, etc. The DIHARD-I dataset also includes a development set and an evaluation set. The durations of the audios in both sets are from 5 to 10 minutes long. The total durations of the audio data are 19.2 hours and 21 hours for the development and evaluation sets, respectively. 
The DER was calculated with a 0-second tolerance collar by following the DIHARD-I challenge \cite{dihard12018}.

\vspace{-.5em}
\subsection{Training data and model configuration}
\vspace{-.5em}

The model training of PET-TSVAD consisted of three stages:
1) pre-train the transformer-based TS-VAD with simulated data and oracle profiles;
2) initialize the PET-TSVAD model with the model from stage 1, and pre-train it with the same simulated data and speaker profiles obtained from clustering;
3) fine-tune the PET-TSVAD model from stage 2 on the development sets of the two tasks described in section~{\ref{eval_data}} independently.

\vspace{-.5em}
\subsubsection{Simulated pre-training data}
\vspace{-.5em}

The pre-training data were generated by multi-talker simulation similar to that used in~\cite{dwang2023_icassp_edaTSvad}.
In the simulation process, we randomly mixed the utterances of 1-10 speakers from VoxCeleb1 \cite{nagrani17VoxCeleb} and VoxCeleb2 \cite{Chung18b} with an overlap ratio up to $30\%$. Noise and reverberations from various sources were applied to the simulated conversations. The signal-to-noise ratio (SNR) was uniformly sampled from 15dB to 40dB when adding additive noises. In total, 21k hours of data were generated, where the duration of each conversation varied from 13 seconds to 8 minutes. 
We chunked the conversation into segments of 60 seconds long,
that was fed into the TS-VAD and PET-TSVAD for training.
For PET-TSVAD training, we extracted the speaker profiles
for each simulated conversation by following the procedure described in 
Section \ref{sec:profile_generation_procedure}.
Specifically, we applied AHC clustering with three different thresholds (0.96, 0.97 and 0.98) and NMESC algorithm on each simulated conversation sample to extract the speaker profiles. The pre-training data was equally divided among four clustering methods plus the setting with oracle profiles. 
The speaker embeddings used for clustering were computed from a Res2Net-based d-vector model \cite{zhou2021resnext}
trained from the VoxCeleb1~\cite{nagrani17VoxCeleb} and VoxCeleb2~\cite{Chung18b}. 
Once the clustering result was obtained, we extracted the speaker profiles
by using the  ResNet-based d-vector model \cite{zhou2019cnn}
trained by VoxCeleb1 and VoxCeleb2 for each detected speaker. 
The oracle speaker profiles 
were extracted based on the ground-truth speaker diarization labels. These oracle profiles were also used to train the baseline transformer-based TS-VAD.

\vspace{-.5em}
\subsubsection{Fine-tuning data}
\vspace{-.5em}

We fine-tuned the models on the real data in the development sets of the VoxConverse and DIHARD-I, respectively. The audio data for each entire session were segmented into 60 seconds for training. The speaker profiles were extracted with clustering algorithms as well. We prepared two sets of configurations regarding the ratio of the clustering algorithms used in the profile extraction, that is shown in Table~\ref{tab:data-ratio-cluster}. 
As shown in the table, 
we applied AHC with various threshold values and NME-SC. We also turned the VAD on and off to further enhance the diversity of the speaker profiles.
To examine the quality of the extracted speaker profiles, we computed the DERs of some of the clustering configurations on the development sets of both VoxConverse and DIHARD-I, which are shown in Table~\ref{tab:data-ratio-der}. It is observed that the AHC with threshold 0.9 and 0.92 achieved significantly better DER than the AHC with threshold 0.97, especially for the VoxConverse development set. From the results of Table~\ref{tab:data-ratio-der}, the overall quality of speaker profile of Config.~2 was higher than that of Config.~1, although Config.~1 included slightly higher percentage of oracle speaker profiles. In addition, the Config.~2 was more diverse than Config.~1 in terms of error patterns in speaker profiles as it included NME-SC and more thresholds for AHC.

\vspace{-.5em}
\subsubsection{Model configuration}
\vspace{-.5em}

The model configuration of PET-TSVAD was the same as the transformer-based TS-VAD \cite{dwang2023_icassp_edaTSvad} except for the pseudo-speaker profile extraction module. We used five 128-dim zero-vectors as input followed by positional encoding and a linear layer to derive the pseudo-speaker profiles that are five 128-dim vectors. These five pseudo-speaker profiles were concatenated with the speaker profiles generated by the first pass diarization, 
and fed to the TS-VAD module of the PET-TSVAD.
The other stream of the input was a sequence of 80-dim log mel-filterbank feature vectors extracted for every 10 ms. The log mel-filterbank features then went through the encoder module to extract the speech embeddings. The encoder module was a 17-layer ResNet, that had the same architecture as the speaker profile extractor except that the final pooling layer was removed. 
The independent speaker detection module consisted of a linear projection layer followed by BLSTM layers. The joint speaker detection module consisted of two blocks of sequence layers where a transformer layer is applied on the speaker-axis followed by a BLSTM layer across the time-axis \cite{dwang2023_icassp_edaTSvad}. The number of parameters of the PET-TSVAD model was 12.47M.

For pre-training the transformer-based TS-VAD model, the parameters were randomly initialized except for the encoder module, which was initialized with the parameters of the profile d-vector extractor. For pre-training the PET-TSVAD, we initialized the parameters with the pre-trained transformer-based TS-VAD model. The batch size during these pre-training stages was set to 64. 100k steps of linear decay learning rate scheduler with 10k steps of warm-up were used with a peak learning rate of 1e-4 and 2e-5 for TS-VAD and PET-TSVAD, respectively. 
For fine-tuning of both TS-VAD and PET-TSVAD models, batch size was set to 32. A fixed learning rate of 1e-5 was used. The total adaptation epochs for TS-VAD and PET-TSVAD were set to 50. Adam optimizer was used for both pre-training and adaptation, with $\beta_1$ and $\beta_2$ equal to 0.9 and 0.98, respectively.

\begin{table}[t]
\centering
\caption{DER (\%) on VoxConverse test set. ``clust-psd-spk'' denotes whether to do clustering on pseudo-speaker profiles' activities across inference chunks. ``sc'' stands for speaker confusion error.}
\label{tab:VoxConverse-result}
  \vspace{0.5mm}
\scalebox{0.84}{
\begin{tabular}{@{}cccccccc@{}}
\toprule
ID  & System                             & Profiles & Clust-psd-spk & DER           & miss & fa  & sc  \\ \midrule
S1  & AHC                             & -        & -             & 7.15          & 4.6  & 1.0 & 1.5 \\
S2  & Xiao et al.\cite{xiao2021microsoft}                               & -        & -             & 6.08          & 2.5  & 2.2 & 1.4 \\
S3  & KrispAI \cite{krisp-arxiv}                          & -        & -             & 4.39          & -    & -   & -   \\ \hdashline[1pt/2pt]\hdashline[0pt/1pt]
S5 & \multirow{2}{*}{TS-VAD} & S1       & -             & 4.59          & 1.9  & 1.3 & 1.3 \\
S6  &                                    & S2       & -             & 4.46          & 1.9  & 1.3 & 1.3 \\ \hdashline[1pt/2pt]\hdashline[0pt/1pt]
S7  & \multirow{4}{*}{PET-TSVAD-1}       & S1       & \checkmark           & 4.53          & 2.1  & 1.2 & 1.2 \\
S8  &                                    & S2       & \checkmark           & 4.45          & 2.1  & 1.3 & 1.1 \\
S9  &                                    & S1       & -            & 4.51          & 2.0  & 1.3 & 1.2 \\
S10  &                                    & S2       & -            & 4.43          & 2.0  & 1.3 & 1.1 \\ \hdashline[1pt/2pt]\hdashline[0pt/1pt]
S11 & \multirow{4}{*}{PET-TSVAD-2}       & S1       & \checkmark           & 4.40          & 2.1  & 1.2 & 1.2 \\
S12 &                                    & S2       & \checkmark           & 4.39          & 2.0  & 1.3 & 1.1 \\
S13 &                                    & S1       & -           & 4.37          & 2.0  & 1.2 & 1.2 \\
S14 &                                    & S2       & -            & \textbf{4.35} & 1.9  & 1.3 & 1.1 \\ \bottomrule
\end{tabular}
}
\end{table}

\vspace{-.5em}
\subsection{Inference settings}
\vspace{-.5em}

We used AHC \cite{xiao2021microsoft} and NME-SC \cite{taejin_eigengap_2019} as the first pass diarization stage for the VoxConverse test set and the DIHARD-I evaluation set, respectively. A speaker profile was then extracted from all the detected non-overlapping speech regions for each speaker. For speakers whose detected speech segments were shorter than 2 seconds, no profile was created and their speech activities were copied to the final diarization results. The test audios were chopped into fixed length chunks (120 seconds) and fed to the transformer-based TS-VAD and the PET-TSVAD for inference. 

\vspace{-.5em}
\subsection{Results}
\vspace{-.5em}

\subsubsection{Results on VoxConverse test set}
\vspace{-.5em}

The results for the VoxConverse test set are presented in Table~\ref{tab:VoxConverse-result}.
Two PET-TSVAD models were trained with two different speaker profile configurations from Table~\ref{tab:data-ratio-cluster}. PET-TSVAD-1 denotes the model trained with the Config. 1
while PET-TSVAD-2 denotes the model trained with the Config. 2 in Table ~\ref{tab:data-ratio-cluster}.
System S1 and S2 are used as the first pass speaker diarization for TS-VAD and PET-TSVAD models. 
Besides the DER results, we list the missing errors (miss), false alarm errors (fa) and speaker confusion errors (sc) for each system. 
The column ``clust-psd-spk'' indicates whether we applied the clustering on the five pseudo-speaker or not. 

From Table~\ref{tab:VoxConverse-result}, we can observe that both PET-TSVAD-1 and PET-TSVAD-2 outperformed the transformer-based TS-VAD when using the same first pass diarization to obtain the speaker profiles. 
In particular, both PET-TSVAD-1 and PET-TSVAD-2  demonstrated consistent improvements on the speaker confusion errors.
In this experiment, PET-TSVAD-2 was slightly better than PET-TSVAD-1.
We speculate that it is because PET-TSVAD-2 had more diversity in target speaker profiles during training. For the rest of the experiments, we will use PET-TSVAD-2 only.

Unexpectedly, we observed that applying the clustering on the five pseudo-speaker profiles slightly but consistently degraded the DER.
We speculate that it was because a detected speaker by pseudo-speaker profiles had a strong correlation with a particular pseudo-speaker profile. It made the consistency of the estimated position of the missing speaker across the inference chunks, and lead the additional clustering unnecessary.

\vspace{-.5em}
\subsubsection{Results on DIHARD-I eval set}
\vspace{-.5em}

Table~\ref{tab:dihard1-result} shows the results for DIHARD-I evaluation set. 
The results show that the proposed PET-TSVAD 
consistently outperformed the TS-VAD baseline. Compared to the transformer-based TS-VAD (S4), the PET-TSVAD-2 (S6) reduced the DER from 27.94\% to 25.88\%, which represents a 7.4\% relative deduction. 
As with the experiment on VoxConverse test set,
we observed better DER by not applying the clustering to the detected speakers corresponding to the pseudo-speaker profiles.

\begin{table}[t]
\centering
\caption{DER (\%) on DIHARD-I eval set.}
\label{tab:dihard1-result}
   \vspace{0.5mm}
\scalebox{0.80}{
\begin{tabular}{@{}cccccccc@{}}
\toprule
ID & System        & Profiles & Clust-psd-spk & DER            & miss & fa   & sc           \\ \midrule
S1 & NME-SC      & -        & -         & 39.87          & 14.0   & 16.7 & 9.2          \\
S2 & Track2 winner \cite{landini_2019_dihard1} & -        & -         & 35.51          & -    & -    & -            \\
S3 & Kwon et al. \cite{kwon21b_interspeech}   & -        & -         & 32.83          & 14.1    & 10.6    & 8.1           \\ \hdashline[1pt/2pt]\hdashline[0pt/1pt]
S4 & TS-VAD        & S1       & -         & 27.94          & 13.3 & 8.0  & 6.6          \\ \hdashline[1pt/2pt]\hdashline[0pt/1pt]
S5 & \multirow{2}{*}{PET-TSVAD-2}      & S1       & \checkmark       & 26.12          & 13.9 & 6.1    & 6.2          \\ 
S6 &     & S1       & -       & \textbf{25.88}          & 15.0 & 5.0    & 5.9          \\ \bottomrule
\end{tabular}}
\end{table}

\vspace{-.5em}
\section{Conclusions}
\vspace{-.5em}

We proposed PET-TSVAD, a transformer-based TS-VAD robust to the speaker profile errors introduced in the first pass diarization. 
On top of the transformer-based TS-VAD that can handle a variable number of speakers, we introduced a set of additional pseudo-speaker profiles to handle speakers undetected during the first pass diarization. 
During training, we used speaker profiles estimated by multiple different clustering algorithms to reduce the mismatch between the training and testing conditions regarding speaker profiles.
We trained the PET-TSVAD model based on PIT to solve the permutation problem
between the output speech activities and the reference labels.
Our experimental results show that PET-TSVAD consistently outperformed the existing TS-VAD models on both the VoxConverse and DIHARD-I datasets.

\section{Acknowledgement}
We thank Yifan Gong, Sunit Sivasankaran, Igor Abramovski, Alon Vinnikov, Tianyan Zhou, Yong Zhao and Gang Liu for the great discussions which bring the valuable insights for our work.

\bibliographystyle{IEEEbib}
\bibliography{my_references,refs}

\end{document}